\documentclass[onecolumn,aps,preprint,showpacs,amsmath,amssymb]{revtex4}
\usepackage{graphicx}
\usepackage{latexsym}
\usepackage{dcolumn} 
\usepackage{stmaryrd} 
 
\usepackage{bm} 

\begin{document}
\newcommand{\eq}{\begin{equation}}                                                                         
\newcommand{\eqe}{\end{equation}}             

\title{Self-similar solutions for the Kardar-Parisi-Zhang interface dynamic equation} 

\author{ I. F. Barna$^a$ and L. M\'aty\'as$^b$}
\address{$^a$ KFKI Atomic Energy Research Institute of the Hungarian Academy 
of Sciences, \\ (KFKI-AEKI), H-1525 Budapest, P.O. Box 49, Hungary, \\ 
 $^b$   Sapientia University, Department of Bioengineering, 
Libert\u{a}tii sq. 1, 530104 Miercurea Ciuc, Romania}

%Email: barnai@sunserv.kfki.hu
\date{\today}

%\maketitle
%%%%%%%%%%%%%%%%%%%%%%%%%%%%%%%%%%%%%%%%%%%%%%%%%%%%%%%%%%%%%%%%%%%%%%%
\begin{abstract} 
In this article we will present a study of the  well-known 
Kardar-Parisi-Zhang(KPZ) model.  
Under certain conditions we have found analytic self-similar solutions for the 
underlying equation. 
The results are strongly related to the error functions. 
One and two spatial dimensions 
are considered with different kind of self-similar Ans\"atze. 
 
\end{abstract}
 
%\draft
\maketitle

%%***********************************************************
\section{Introduction}

Growth of patterns in clusters and sodification fronts are a challenging problem from a long time. 
Basic knowledge of the roughness of growing crystalline facets has obvious technical applications \cite{konyv}. 
The simplest nonlinear generalization of the ubiquitous diffusion equation is the the so called Kardar-Parisi-Zhang \cite{kpz} model obtained from Langevin equation 
\eq
\frac{\partial h}{\partial t} = \nu {\bf{\nabla}}^2 h + \frac{\lambda}{2}({\bf{\nabla}} h)^2 + \eta({\bf{x}},t)
\label{kpz}
\eqe
where h stands for the level of the local profile. 
The first term on the right hand side describes relaxation of the interface be a surface tension which prefers a smooth surface.  
The second term is the lowest-order nonlinear term that can appear in the
surface growth equation justified with the Eden model and originates from the
tendency of the surface to locally grow normal to itself and is
non-equilibrium in origin. The last term is a Langevin noise to mimic the
stochastic nature of any growth process and has a Gaussian distribution
usually.  
 The right hand side eventually may contain a constant term corresponding
  to a driving force that generates
  an  increase of the surface and which will be discussed later. 
There are numerous studies available about the KPZ equation in the literature in the last fifteen years. 
Without completeness we mention some of them. 
Hwa and Frey \cite{hwa1} investigated the KPZ model with the help of the self-consistent mode-coupling method and with  renormalisation group-theory which is an exhaustive and sophisticated method which uses Green's functions. 
They considered various dynamical scaling form of $C(x,t) = x^{-2 \chi} C(bx,b^zt)$ for the correlation function. 
L\"assig shows how the KPZ model can be derived and investigated with field theoretical approach \cite{lass}. 
In a topical review paper Kriecherbauer and Krug \cite{krug} derives the KPZ model from hydrodynamical conservation equations with a general current density relation.  

 One may find similar models, which may lead to similar equations to the
  one presented above, modeling the interface
  growth of bacterial colonies \cite{Matsushita}. 
 There is a more general interface growing model based on the so-called 
Kuramoto-Sivashinsky   \cite{kuram} equation which is basically the KPZ model with and extra $ -\nabla^4 h$ term on the right hand side of (1). 
In the following we will not consider this model, because the self-similar Ansatz cannot be applied to this  equation. 
%%************************************************
\section{Method for solution}
To obtain analytic solutions of the KPZ equation in the following  we will use the self-similar Ansatz which 
can be found in \cite{sedov,barenb,zeld} 
\eq 
T(x,t)=t^{-\alpha}f\left(\frac{x}{t^\beta}\right):=t^{-\alpha}f(\omega), 
\label{self}
\eqe 
where $T(x,t)$ can be an arbitrary variable of a PDE and $t$ means time and $x$ means spatial 
dependence.
The similarity exponents $\alpha$ and $\beta$ are of primary physical importance since $\alpha$  
represents the rate of decay of the magnitude $T(x,t)$, while $\beta$  is the rate of spread 
(or contraction if  $\beta<0$ ) of the space distribution as time goes on.
The most powerful result of this Ansatz is the fundamental or 
Gaussian solution of the Fourier heat conduction equation (or for Fick's
diffusion equation) with $\alpha =\beta = 1/2$.  (Exponents with a value of one half in another 
language means a typical 
random walk process.)
These solutions are visualized on figure 1. for time-points $t_1<t_2$. 
Solutions with integer exponents are called self-similar solutions of the first kind, non-integer exponents 
mean self-similar solutions of the second kind.
The existence of self-similar solutions exclude the existence any kind of characteristic time scale as well.  \\  
Applicability of this Ansatz is quite wide and comes up in various 
mechanical systems \cite{sedov,barenb,zeld}, transport phenomena like heat conduction \cite{ barn} 
or even the three dimensional Navier-Stokes equation \cite{barna2}. 
%%%%%%%%%%%%%%%%%%%%%%%%%%%%%%%%%%%%%%%%%%%%%%%%%%%%%
\begin{figure} 
% \vspace*{1.0cm} 
 %\hspace*{0.4cm} 
\scalebox{0.4}{
\rotatebox{0}{\includegraphics{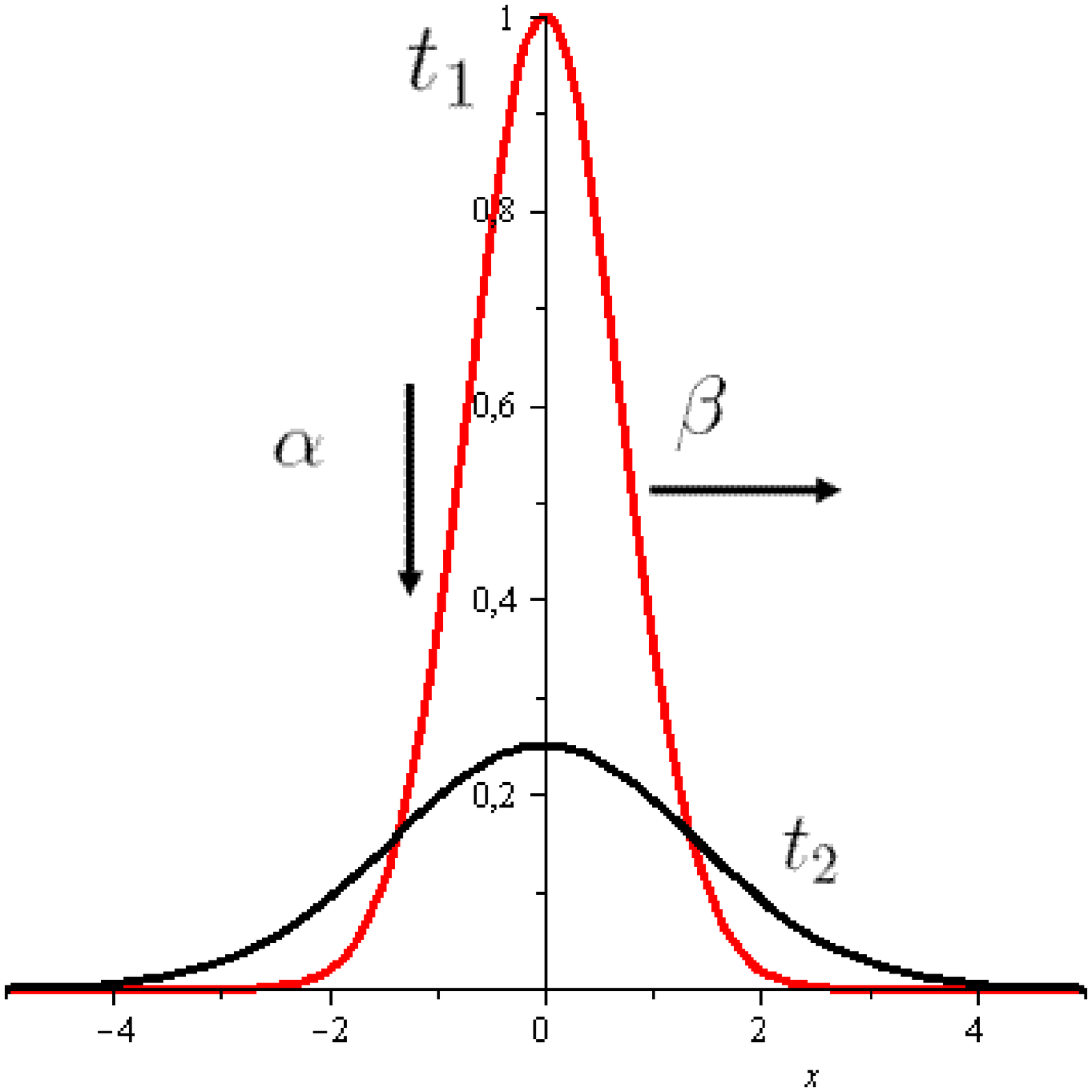}}}
\vspace*{-1.4cm}
\caption{A self-similar solution of Eq. (\ref{self}) for $t_1<t_2$.
The presented curves are Gaussians for regular heat conduction.}	
\label{egyes}       % Give a unique label
\end{figure}
%%%%%%%%%%%%%%%%%%%%%%%%%%%%%%%%%%%%%%%%%%%%%%%%%%

%%%%%%%%%%%%%%%%%%%%%%%%%%%%%%%%%%%%%%%%%%%%%%%%%%%%%
\begin{figure}  
%\vspace*{1.0cm}
%\hspace*{0.4cm}
\scalebox{0.35}{
\rotatebox{0}{\includegraphics{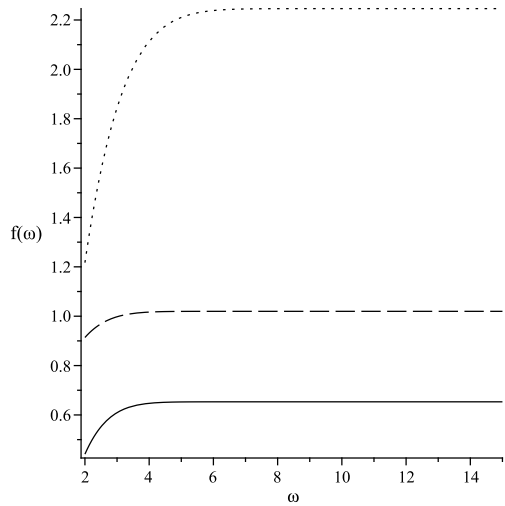}}}
%\vspace*{0.4cm}
\caption{The self similar solution Eq. ( \ref{sol})  for $c_1=c_2 =1$. Solid line is for $\lambda = \nu =1$, dashed line is for 
$\nu =1$ and $\lambda =2$ and the dotted line represents $\nu =2$ and $\lambda =1$.}	
\label{kettes}       % Give a unique label
\end{figure}
%%%%%%%%%%%%%%%%%%%%%%%%%%%%%%%%%%%%%%%%%%%%%%%%%%%%%%%%%%%%
In the following we will consider one spatial dependence of the KPZ equation. 
Calculating the first time and space derivative of the Ansatz (\ref{self}) and pluging back to (\ref{kpz}) 
(first we consider no noise term $\eta(x,t)=0$) we get the following constrains for the exponents: 
$\alpha = 0$ and $\beta = 1/2$. 
The remaining non-linear ordinary differential equation(ODE) reads
\eq
\nu f''(\omega) + f'(\omega)\left [\frac{\omega}{2}+ \frac{\lambda}{2} f'(\omega)\right] = 0.  
\label{ode}
\eqe
The result is proportional with the logarithm of the error function 
\eq
f(\omega) = \frac{2\nu ln \left( \frac{\lambda c_1 \sqrt{\pi\nu} \> \> erf[\omega/(2\sqrt{\nu})] + c_2}{2\nu}  \right)  } {\lambda}
\label{sol} 
\eqe
where $erf$ is the error function  \cite{abr} and $c_1$ and $c_2$ are integration constants. 
The role of $c_2$ is just a shift of the solution. From physical reasons $\nu$ the surface tension should be larger than zero. 
From analysis of the solution Eq. (\ref{sol}) the value of $
\lambda$ should be positive as well. 
Figure 2. presents two solutions with $c_1 = c_2 = 1$ and for three different $
\lambda $ and $\nu$ combinations. 
Note, that all solution has the same simple qualitative behavior, a ramp-up and a converged plateau.
Figure 3. shows a three dimensional x,t dependent solution of the original equation for $c_1=c_2=\lambda= \nu =1$ 
The function has a similar structure like Fig. 2. a ramp-up and a convergent plateau. 

Let's consider some analytic noise term for the 1+1 dimensional case.
From the self-similar Ansatz (2) it is obvious that the noise term $\eta(x,t)$ should be a function of  
$\omega = x/t^{\beta}$, on the other side $\eta(x,t)$ should be a distribution function as well. 
Therefore we first tried the following noise terms: $exp(-\omega)$ and $exp(-\omega^2)$. Unfortunately, no analytic 
results could be found in any closed form. For the $1/(1+\omega^2)$ (Lorenzian) noise term we got analytic results 
which can be expresses in sophisticated terms of Heun C special functions which we not mention here. 
A second analytic result was found for the $1/\omega^2$ noise term which can be evaluated with a tedious combination of 
the Whitakker M and Whitakker W functions which we also skip here.     

If the surface growth is fed by some mechanism with a driving force, like the
constant $\nu$ term that may appear in addition on the r.h.s. of Eq. 1.
and which may be suggested by the work of \cite{vicsek},
then other analytical solutions become available which can be expressed via Kummer M and Kummer U functions. 
If the constant driving term in its magnitude is not $\nu$ (not equal with the parameter of the
Laplacean term) 
but a general $\epsilon$, the solutions are even more complicated. 
There is no general closed form for this case where all the three parameters are free. 
If we fix $\lambda = \nu  =1$ than three independent cases are available. 
For $\ 0 < \epsilon < 1$ the solution is like 
in Eq. 4. If $\epsilon  = 2$ the solutions contains a complex part, however for large $(\epsilon =10)$      
the solution is evaluated via the combination of  hypergeometric functions. 
Note, that all these solutions contain two integrations constants $c_1, c_2$ which have non-linear dependence
in the solutions.  
%%%%%%%%%%%%%%%%%%%%%%%%%%%%%%%%%%
 \begin{figure}   
%\vspace*{1.0cm}
%\hspace*{0.4cm}
\scalebox{0.35}{
\rotatebox{0}{\includegraphics{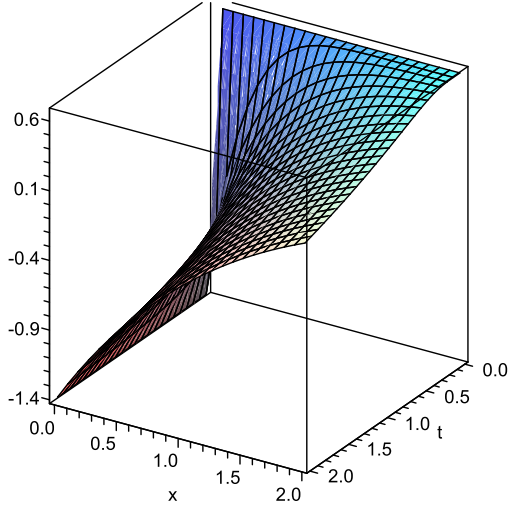}}}
%\fbvspace*{0.4cm}
\caption{The self-similar solution of the original Equation for $c_1 = c_2 = \lambda = \nu =1$.}	
\label{harmas}       % Give a unique label
\end{figure}
%%%%%%%%%%%%%%%%%%%%%%%%%%%%%%%%%

At last we investigate the two spatial dependent cases where we generalize the self-similar 
Ansatz. In our former studies it came out clearly that the self-similar Ansatz of (2) can be generalised in many 
ways \cite{barn}/2: 
\eq 
h(x,y,t)=t^{-\alpha}f\left(\frac{F[x,y]}{t^\beta}\right):=t^{-\alpha}f(\omega), 
\label{2d1} 
\eqe
where 
$F[x,y]$ can be understood as a parametrization of a 2 dimensional curve. 
For $F[x,y]$ we considered various functions like the most important  $\sqrt{x^2 + y^2} = a$ which can be interpreted 
as the usual distance or the $L^2$  Euclidean norm. Unfortunately, now for the recent KPZ model only the linear case $0=y-ax-b$ is available giving us the following ODE without any contradiction 
 
\eq
\nu(a^2+1) f''(\omega) + f'(\omega)\left [\frac{\omega}{2}+ \frac{\lambda}{2} f'(\omega)(a^2-1) \right] = 0.  
\eqe
For the exponents the $\alpha =0$ and $\beta = 1/2$ constrains are still valid. 
The general analytical solution is the following: 
\begin{eqnarray}
f(\omega) =  \frac{1}{\lambda(a^2-1)} \left(ln \left[ \frac{1}{4} \frac{1}{\nu(a^2+1)}
\left(  \lambda^2 \left\{ c_1 \sqrt{\pi} erf \left(\frac{1}{2} \sqrt{\frac{1}{\nu(a^2+1)}} \omega \right) a^2 - 
  \right.  \right.  \right.  \right.  
 \nonumber \\
 \left. \left. \left.  \left. 
 c_1 \sqrt{\pi} erf \left(\frac{1}{2} \sqrt{\frac{1}{\nu(a^2+1)}} \omega\right) +  
c_2 \sqrt{\frac{1}{\nu(a^2+1)}} a^2 -
c_2 \sqrt{\frac{1}{\nu(a^2+1)}}  
\right\}^2 \right) \right] \nu(a^2+1)\right).
\end{eqnarray}
Note, that this solution is very similar to the one dimensional one (4) 
there is a ramp-on and a plateau for $0<\omega$ which means positive time and positive 
interface widths these are physical constraints of a real solution. Figure 4 presents a particular 
solution for the $\lambda = \nu = c_1 = c_2 =1,  a=2$ parameters.

%%%%%%%%%%%%%%%%%%%%%%%%%%%%
 \begin{figure}   
%\vspace*{1.0cm}
%\hspace*{0.4cm}
\scalebox{0.35}{
\rotatebox{0}{\includegraphics{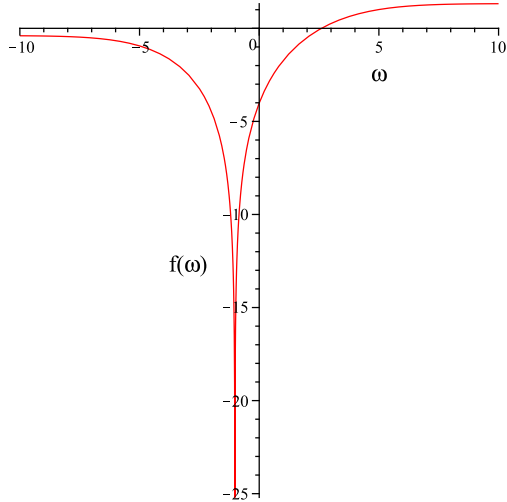}}}
%\fbvspace*{0.4cm}
\caption{Solution of Eq. 7 for $c_1 = c_2 = \lambda = \nu =1, a=2$.}	
\label{harmas}       % Give a unique label
\end{figure}
%%%%%%%%%%%%%%%%%%%%%%%%% 

%%**************************************************************
\section{Conclusions}
In summary we can say, that with an appropriate change of 
variables applying the self-similar Ansatz one may obtain 
analytic solution for the KPZ equation for one or two dimension sometimes even with some noise term.  
%%%%%%%%%%%%%%%%%%%%%%%%%%%%%%%%%%%%%%%%%%%%%%%%%%%%%%%%%%%%%%%%%%%%%%                                    

%%%%%%%%%%%%%%%%%%%%%%%%%%%%%%%%%%%%%%%%%%%%%%%%%%%%%%%%%%%%%%%%%%%%%%%                                    
%\end{multicols}
%%%%%%%%%%%%%%%%%%%%%%%%%%%%%%%%%%%%%%%%%%%%%%%%%%%%%%%%%%%%%%%%%%%%%%%                                    

\end{document}